\documentclass[twocolumn,showpacs,preprintnumbers,amsmath,amssymb]{revtex4}

\usepackage{graphicx}
\usepackage{dcolumn}
\usepackage{bm}

\newcommand{\chiup}{\raisebox{0.2em}{$\chi$}}

\begin{document}

\preprint{}

\title{Electromagnetic Magic: The Relativistically Rotating Disk}

\author{D. Lynden-Bell$^{1,2}$}
\affiliation{
$^1$The Institute of Astronomy, The Observatories,\\
Madingley Road, Cambridge, CB3 0HA, UK\\
$^2$Institute of Advanced Study, Princeton NJ
}%

\date{\today}

\begin{abstract}
A closed form analytic solution is found for the electromagnetic field
of the charged uniformly rotating conducting disk for all values of
the tip speed $v$ up to $c$. For $v=c$ it becomes the Magic field of the
Kerr-Newman black hole with $G$ set to zero.

The field energy, field angular momentum and gyromagnetic ratio are
calculated and compared with those of the electron.

A new mathematical expression that sums products of 3 Legendre
functions each of a different argument, is demonstrated.
\end{abstract}


\maketitle

\section{\label{sec:introduction}Introduction}
Away from sources we write the electromagnetic field
${\bf{F}}={\bf{E}}+i{\bf{B}}=-\nabla\Psi$ where $\Psi=\Phi +i\chiup$ is
a complex potential satisfying $\nabla^2\Psi=0$. The simplest solution
is the real one $\Psi=q/r$ but if we make $q$ complex we introduce a
magnetic monopole which is unacceptable. Moving the charge to
$\tilde{\bf{a}}$ we have a potential $q/\sqrt{(\bf{r-\tilde{a}})^2}$
which corresponds to a charge at $\tilde{\bf{a}}$ when
$\tilde{\bf{a}}$ is real, but we now consider $\tilde{\bf{a}}$ to be
complex ${\bf{r}}_1+i\bf{a}$. Without loss of generality we may move the
origin to $\bf{r}_1$ and orient the $z$ axis along $\bf{a}$. If we
then set $z=r\cos\theta=r\mu$ we find
\begin{eqnarray}\label{1}
\Psi=q/\sqrt{({\bf{r}}-i{\bf{a}})^2}&=&q/(r^2-2iar\mu-a^2)^{1/2}\\
&=&\frac{q}{r}\sum\left(\frac{ia}{r}\right)^nP_n(\mu)~~ {\rm
for }~r \geqslant a~. \nonumber
\end{eqnarray}
$\Psi$ has a singular ring at $\mu=0$ and $r=a$. To specify $\Psi$
properly we need to say which square root is to be taken, but the
expansion above for $r\geqslant a$ is required to make everything
regular at infinity and the simplest cut is across the circle defined
by the ring singularity. With such a cut analytic continuation defines
$\Psi$ everywhere. The `magic' field is given by
\begin{equation}\label{2}
{\bf{F=E}}+i{\bf{B}}=-\nabla\Psi
=q\frac{{\bf{r}}-i{\bf{a}}}{[({\bf{r}}-i{\bf{a}})^2]^{3/2}} 
\end{equation}
By construction $\nabla^2\Psi=0$ except on the singular ring and the
cut, so the sources of the field lie there. The expansion (\ref{1})
shows that the field has charge $q$, no magnetic monopole but a
magnetic dipole $qa$. Indeed the electric field has only even
multipoles and the magnetic field has only odd ones. On
$z=0,~{\bf{a.r}}=0$ so the denominator of (\ref{2}) is real if
$r\geqslant a$ and pure imaginary for $r < a$.

Thus close to the plane of symmetry 
$${\bf{E}}=-q{\bf{a}}/(a^2-r^2)^{3/2}~,
$$
and ${\bf{B}}=-q{\bf{r}}/(a^2-r^2)^{3/2}$ for $z>0$ and $r<a$. Below,
where $z<0$ these fields are reversed. Notice that $\bf{E}$ is
orthogonal to the disk so the disk is an equipotential and indeed its
potential is zero (earthed) as may be seen by taking the real part of
(\ref{1}) on $z=0$ with $r^2<a^2$. The magnetic field lies parallel to
the radius below the disk and anti-parallel above. It does not cross
the disk except at the singular ring. For $r>a$ and $z=0$, $\bf{B}$
points downwards everywhere. Thus every field line returns to the
upper hemisphere through the singular ring.

From the electric field one readily calculates the surface density of
charge on the disk (summing both sides)
\begin{equation} \label{3}
\sigma=-\frac{qa}{2\pi (a^2-r^2)^{3/2}}
\end{equation}
Notice that this is of the opposite sign to the total $q$ which is
rectified only at the singular ring. Indeed the total charge on the
cut within $r=R$ is
$Q (<R) = - q\left(\frac{a-\sqrt{a^2-R^2}}{\sqrt{a^2-R^2}}\right)$ 
for $R<a$ which diverges with a negative sign as $R$ approaches $a$
but $Q(<R)=+q$ for $R>a$ so the negative infinity is now replaced by a
finite positive result. Likewise the surface current in the cut is
\begin{equation}\label{4}
J_\phi = -\frac{q}{2\pi}~\frac{r}{(a^2-r^2)^{3/2}}
\end{equation}
which is precisely what we would get if the above charge density were
rotating rigidly with angular velocity $\Omega=c/a$. The total current
within $R<a$ is $Q(<R)\Omega/(2\pi c)$. Again the magnetic effects of
this current are overwhelmed by the current around the singular ring
which is of opposite sign. Notice that the disk is not crossed by a
magnetic field line saving at the singular edge ring itself. In this
respect the disk acts like a superconductor displaying the Meissner
effect. In the Black Hole context this phenomenon was noted by
Bi\u{c}\'{a}k \& Jani\u{s} [1] (see also Bi\u{c}\'{a}k \& Ledvinka
[2]).

I now list other properties of this `Magic' electromagnetic
field. (For proofs see Lynden-Bell [3])

\begin{enumerate}
\item
Relativistic Invariants
$F^2=E^2-B^2+2i{\bf{E.B}}=q^2/({\bf{r}}-i{\bf{a}})^4$. 

\item
$E^2=B^2$ only on two spheres of radius $\sqrt{2}a$ centred on
${\bf{r}}=\pm {\bf{a}}$. They meet on the singular ring.

\item
${\bf{E.B}}=0$ on the sphere $r=a$ and also on the plane $z=0$.

\item
The field energy density ${\bf{F.F^*}}/8\pi =
[q^2/(8\pi)](r^2+a^2)/|({\bf{r}} -i{\bf{a}})^2|^3$. This diverges like
$(r-a)^{-3}$ when $z=0$ near $r=a$.

\item
${\bf{F^*}} \times
{\bf{F}}=2iq{\bf{E}}\times{\bf{B}}=2iq^2{\bf{a}}
\times{\bf{r}}/|({\bf{r}}-i{\bf{a}})^2|^3$ likewise diverges.

\item
The above is related to the Poynting vector which shows that the
energy density flows around the axis with a velocity
$\bm{\Omega}\times{\bf{r}}$ where $\bm{\Omega}$ is constant on spheres
$r=$const with $\bm{\Omega}(r)=\frac{2{\bf{a}}c}{r^2+a^2}$.

\item
Landau \& Lifshitz [4] show that in the frame that moves with the velocity
$c{\bf{V}}$ where ${\bf{V}}/(1+V^2)={\bf{E}}\times{\bf{B}}/(E^2+B^2)$,
the transformed fields ${\bf{E}}^\prime \& {\bf{B}}^\prime$ are
parallel. Gair proved the theorem that these velocities are a uniform
rotation of each spheroid confocal with the disk at the rate
$\bm{\Omega}={\bf{a}}c/(\tilde{r}^2+a^2)$ (for $\tilde{r}$ see section
2).

\item
In the corotating frame of those spheroids ${\bf{E}}^\prime$ and
${\bf{B}}^\prime$ are both perpendicular to the spheroid on which they
lie.

\item
The total field energy and the total angular momentum in the field
both diverge due to their divergence at the singular ring.

\end{enumerate}

More remarkable properties still come from the separability of both
wave equations and equations of motion of any relativistic charged
particle in this field.

\begin{enumerate}
\item
The Hamilton-Jacobi equation separates

\item
The Klein Gordon equation separates

\item
The Dirac equation separates

\item
The Schr\"odinger equation separates if only the real part of the
field is included (but not when the magnetic part is added although
the reverse is true for the Klein Gordon equation).

\end{enumerate}

All the above really stem from remarkable investigations in the
separability of wave equations around black holes by Carter [5],[6],
Teukolsky [7], Chandrasekhar [8] and Page [9]. In the simpler flat
space case see Lynden-Bell Paper I [10].

Finally the relationship with black holes and indeed the original
discovery of the field in General Relativity is due to Newman {\it et
al.} [11] who generalised the Kerr metric of a spinning black hole to
include charge. If in his solution one puts Newton's $G\equiv 0$ one
obtains an electromagnetic field in flat space which is the above
`Magic' field [12], [13], [14].

The present investigation is aimed at providing a set of
electromagnetic fields in flat space that keep some of the remarkable
properties of the Magic field but give finite answers for total field
energy, total field angular momentum etc. We aim to find fields which
give the Magic field as a limiting case but whose other members give
finite answers. Some spice is added to the investigation by Carter's
[5] remark that all Kerr Metrics have the same gyromagnetic ratio, 2,
as the Dirac electron, and Pfister \& King [15] propose that this may
have a deeper significance for general relativity. Indeed it holds for
{\bf all} the conformastationary metrics, the statement by us [16]
that the gyromagnetic ratio was one, was based on erroneously missing
out the factor 2 in its definition. Our disk [16] remains the only
known conformastationary interior solution to Einstein's equation.

Recently [17] we discussed the remarkable behaviour of charges on a
relativistically rotating conducting sphere. The changes in the field
correspond to addition of more and more of the Magic field as the
rotation increases. However even when $v=c$ the distribution on the
sphere never becomes the Magic field: that discussion leads to the
suggestion that the Magic field may be the field of the rapidly
rotating charged conducting plate with a tip speed of $c$. The facts
that the component of ${\bf{E}}+{\bf{V}}\times{\bf{B}}$ in the plate
is zero and that the current is due to the convected charge strongly
suggest that this is the limit, as $Vc$ the tip speed tends to $c$, of
the charged rotating conducting plate. This paper is devoted to a
discussion of the whole sequence of fields of such a plate when it
rotates at any tipspeed up to $c$. The field lines of the Magic field
are shown in Figure 1, which was generated by J. Gair.

\begin{figure*}
\includegraphics{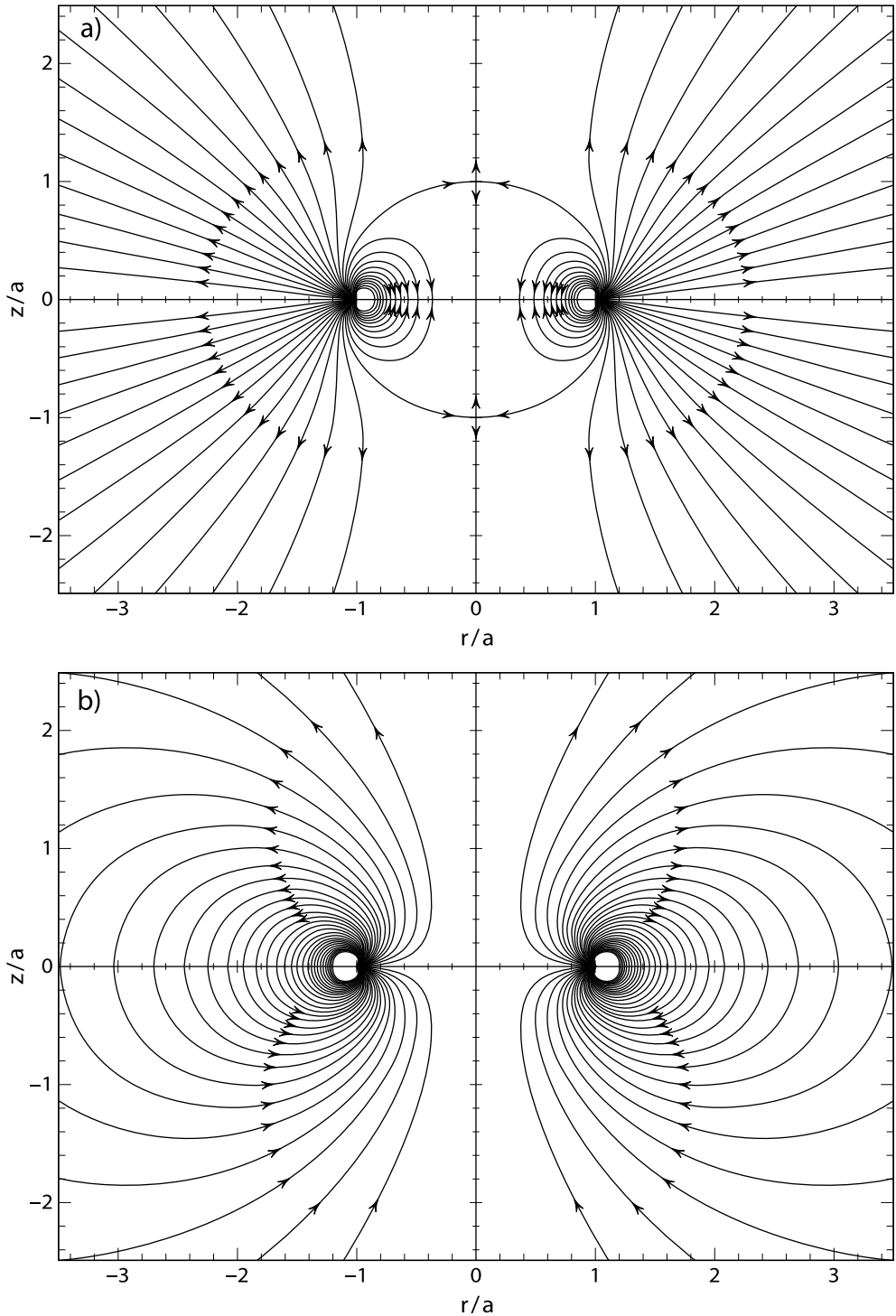}
\caption{\label{fig:figure} (a) Electric lines of force for the
Kerr-Newman potential. The density of lines reflects the strength of
the field. \\(b) Magnetic lines of force for the Kerr-Newman potential. The
density of lines reflects the strength of the field.}
\end{figure*}

\section{Maxwell's Equations for the Rotating Disk}

We treat the problem using oblate spheroidal coordinates confocal with
the ring that forms the edge of the disk. Actually we shall solve the
problem as the limit of a highly flattened spheroid as its minor axis
$\tilde{b}$ tends to zero; this allows us to make direct comparisons
with our treatment of the rotating charged conducting sphere. The
equation

\begin{equation}\label{5}
\frac{x^2+y^2}{\tau+a^2}+\frac{z^2}{\tau}=1
\end{equation}
defines a set of confocal oblate spheroids each with a definite
positive value of $\tau$. When $\tau=0$ it defines a disk
$z=0,~x^2+y^2\leqslant a^2$. When $\tau$ is negative it defines a set
of hyperboloids of revolution (of one sheet) each of which is
orthogonal to the oblate spheroids. For $\tau$ positive we write it as
$\tilde{r}^2$ and for $\tau$ negative we write it as $-a^2\cos
^2\vartheta$ or more often as $-a^2\tilde{\mu}^2$. The spheroidal
coordinates $\tilde{r},\vartheta$ are supplemented by the usual
azimuthal angle $\phi$ to give the metric
\begin{eqnarray} \label{6}
ds^2&=&\frac{\tilde{r}^2+a^2\tilde{\mu}^2}{\tilde{r}^2+a^2}d\tilde{r}^2
+\frac{(\tilde{r}^2+a^2\tilde{\mu}^2)}{1-\tilde{\mu}^2}d\bar{\mu}^2
\nonumber\\&+&(1-\tilde{\mu}^2)(\tilde{r}^2+a^2)d\phi^2 \\ &=&
dr^2+r^2d\theta^2+r^2\sin^2\theta d\phi^2 \nonumber\\ &=&
dR^2+R^2d\phi^2+dz^2 \nonumber
\end{eqnarray}
where $\theta$ is the spherical polar coordinate which should be
distinguished from the spheroidal coordinate $\vartheta$ although they
are equal at $\infty$. We consider the surface of the spheroid
$\tilde{r}=\tilde{b}$ to be the rotating conductor.

On $\tilde{r}=0$, the disk, $\partial\chiup/\partial\tilde{r}$ changes
sign which corresponds to the continuity of the normal magnetic flux,
and $\Phi$ is continuous which corresponds to the surface component of
the electric field being continuous.

The other requirements are that the total charge must be $q$, so $\Phi
\rightarrow q/r$ at $\infty$ and
${\bf{E}}+\frac{1}{c}(\bm{\Omega}\times{\bf{r}}) \times {\bf{B}}$ must
have no component along the disk's surface so that the Lorentz force
component vanishes. Of course the discontinuity of the surface
component of $-\nabla\chiup$ corresponds to the surface current $4\pi
J_\phi$ and the discontinuity in the normal component of $-\nabla\Phi$
gives the surface density of charge $4\pi\sigma$. In the steady state
with an imperfect conductor the currents are just those caused by the
rotation of the charge.

\begin{equation} \label{7}
{\bf{J}} = \frac{1}{c}(\bm{\Omega} \times {\bf{r}})\sigma
\end{equation}

That completes the physical specification of the problem. Now we turn
these physical requirements into mathematics.

On the disk
\begin{eqnarray*}
\mathcal{I}m(-\partial\Psi/\partial R)&=&B_R=2\pi J_\phi\\
\mathcal{R}e(-\partial\Psi/\partial z)&=&2\pi \sigma
\end{eqnarray*}
and from the Lorentz force component
\begin{equation*}
\mathcal{R}e(-\partial\Psi/\partial R) + (\Omega
R/c)\mathcal{I}m(-\partial\Psi/\partial z)=0
\end{equation*}
while (\ref{7}) gives 
$$
\mathcal{I}m(-\partial\Psi/\partial R)-(\Omega
R/c)\mathcal{R}e(-\partial\Psi/\partial z)=0
$$
Hence on the disk $R<a$, $z=0$
\begin{equation} \label{8}
\partial\Psi/\partial R = i(\Omega R/c)\partial\Psi/\partial z
\end{equation}

Now in spheroidal coordinates the general solution of Laplace's
equation which tends to $q/r$ at $\infty$ is 
\begin{equation}  \label{9}
\Psi=\frac{q}{ia} \sum^\infty_0
\Psi_l~P_l(\tilde{\mu})Q_l\left(\frac{\tilde{r}}{ia}\right)
\end{equation}
$P_l(\tilde{\mu})$ are the Legendre Polynomicals and the
$Q_l\left(\frac{\tilde{r}}{ia}\right)$ are the Legendre functions here
expressed as functions of an imaginary argument. The
$Q_{2l}\left(\frac{\tilde{r}}{ia}\right)$ are imaginary while the
$Q_{2l+1}\left(\frac{\tilde{r}}{ia}\right)$ are real. The $Q_l(z)$ obey the
recurrence relations $(l+1)Q_{l+1}(z)=(2l+1)zQ_l(z)-lQ_{l-1}(z)$ and
$(z^2-1)Q_l^\prime(z)=lzQ_l(z)-lQ_{l-1}(z)$.

Hereafter whenever we refer to $Q_l$ or $Q_l^\prime$ without explicit
mention of its argument we shall mean $Q_l(0)$ and $Q_l^\prime(0)$
i.e. the limit of $Q_l\left(\frac{\tilde{b}}{ia}\right)$ as
$\tilde{b}\rightarrow 0$ etc. We then have
\begin{equation} \label{10}
Q_{l-1}=-[(l+1)/l]Q_{l+1}~~{\rm and}~~Q_l^\prime =lQ_{l-1}=-(l+1)Q_{l+1}
\end{equation}

\begin{eqnarray*}
Q_0(z)=\frac{1}{2}\ln\left(\frac{z-1}{z+1}\right)&\rightarrow &
-\frac{i\pi}{2}~~{\rm as}~~z\rightarrow\frac{b}{ia}\rightarrow 0\\
&\rightarrow & ia/\tilde{r}~~{\rm
as}~~z=\frac{\tilde{r}}{ia}\rightarrow\infty\\
 Q_1(z)=\frac{z}{2}\ln
\left(\frac{z+1}{z-1}\right)-1&\rightarrow &-1~~{\rm as}~~z
\rightarrow 0
\end{eqnarray*}

The condition that $\Psi \rightarrow q/r$ at $\infty$ becomes
$\Psi_0=1$. Rewriting our boundary condition (\ref{8}) in spheroidal
coordinates
\begin{equation*}
-\frac{\sqrt{1-\tilde{\mu}^2}}{a|\tilde{\mu}|}~
\frac{\partial\Psi}{\partial\tilde{\mu}} = \frac{i\Omega
a}{c}~\frac{\sqrt{1-\tilde{\mu}^2}}{|\tilde{\mu}|}~
\frac{\partial\Psi}{\partial\tilde{r}}~~~{\rm on}~~\tilde{r}=0
\end{equation*}
Thus from (\ref{9}) setting $\Omega a/c=\omega$
\begin{equation} \label{11}
\sum^\infty_0\Psi_l~P_l^\prime(\tilde{\mu}) Q_l=
-\omega\sum^\infty_0\Psi_lP_l(\tilde{\mu})Q^\prime_l
\end{equation}

Now as well as their usual orthogonality relation the Legendre
polynomicals $P_l(\mu)$ have the property
\begin{equation*}
\int^{+1}_{-1}\!P^{\prime}_{l+1}(\mu)~P_{m}(\mu)~d \mu = 
\begin{cases}
2~{\rm for}~l\geqslant m,~(l+m)~{\rm even}\\
0~{\rm for}~l<m~{\rm or}~(l+m)~{\rm odd}
\end{cases}
\end{equation*}
Multiplying (\ref{11}) by $\frac{1}{2}P_{2m}(\tilde{\mu})$ and
integrating $-1$ to $+1$
\begin{equation} \label{12}
\sum^\infty_m
(4l+3)\eta_{2l+1}~Q_{2l+1}=-\omega\eta_{2m}~Q^\prime_{2m}=-
\omega\eta_{2m}2m Q_{2m-1}
\end{equation}
where $\eta_m=\Psi_m/(2m+1)$. Subtracting this with $(m+1)$ written
for $m$
\begin{eqnarray*}
(4m+3)\eta_{2m+1}Q_{2m+1}&=&\omega[\eta_{2m+2}(2m+2)Q_{2m+1}\\
&+&\eta_{2m}(2m+1)Q_{2m+1}]
\end{eqnarray*}
where we used (\ref{10}) on the last term. Thus
\begin{equation} \label{13}
(4m+3)\eta_{2m+1}=\omega\left[ (2m+2)\eta_{2m+2} +(2m+1)\eta_{2m}\right]
\end{equation}
similarly multiplying (\ref{11}) by $\frac{1}{2}P_{2m+1}(\tilde{\mu})$
and integrating etc
\begin{equation} \label{14}
(4m+5)\eta_{2m+2}=\omega\left[(2m+3)\eta_{2m+3}+(2m+2)\eta_{2m+1}\right]
\end{equation}
(\ref{13}) \& (\ref{14}) can be recombined into a single formula true
for all $m$
\begin{equation} \label{15}
(2m+1)\eta_m=\omega\left[(m+1)\eta_{m+1}+m\eta_{m-1}\right]~.
\end{equation}

When the Magic potential is rewritten in terms of the spheroidal
coordinates $\tilde{r}$ and $\tilde{\mu}$ it takes a remarkably simple
form and it is interesting to resolve it into Legendre polynomials of
$\tilde{\mu}$
\begin{eqnarray*}
\Psi&=&q/\sqrt{({\bf{r}}-i{\bf{a}})^2}=q/(\tilde{r}-ia\tilde{\mu})\\
&=&
\left(\frac{q}{ia}\right) \sum^\infty_0
M_l\left(\frac{\tilde{r}}{ia}\right) P_l(\tilde{\mu})
\end{eqnarray*}
We multiply both sides by $P_n(\tilde{\mu})$ and integrate $-1$ to
$+1$. By Abramowitz \& Stegan 8.8.3 $\frac{1}{2}\int^{+1}_{-1}
\frac{P_n(\mu)}{(z-\mu)} d\mu = Q_n(z)$ so we find
$M_l\left(\frac{\tilde{r}}{ia}\right)=(2l+1)Q_l
\left(\frac{\tilde{r}}{ia}\right)$ and in terms of the expansion
(\ref{9}) $\Psi_l=2l+1$ so $\eta_l=1$ for the Magic field. We see at
once that this satisfies (\ref{15}) with $\omega$ set equal to 1 so
that the Magic field is as expected the limit of the uniformly
rotating disk when $v\rightarrow c$. To solve our problem for a
general rotation rate we use the generating function method developed
in the appendix of Paper III [17].

We multiply equation (\ref{15}) by $U^m$ and sum from $m=1$ to
$\infty$. Defining $\eta(U)=\sum\limits^\infty_0\eta_mU^m$ and remembering
that $\eta_0=\Phi_0=1$ we find that
\begin{equation*}
2U\eta^\prime+\eta-1=\omega(\eta^\prime-\eta_1+U^2\eta^\prime+\eta)
\end{equation*}

We rewrite this in terms of the roots $u\leqslant 1$ and $1/u$ of the
quadratic $t^2-\frac{2}{\omega}t+1=0$ so
$\frac{1}{\omega}=\frac{1}{2}(u^{-1}+u)$. [Notice that the quadratic
and $u$ differ from those of paper III in which we discussed the
sphere rather than the disk]
\begin{equation} \label{16}
(U-u)(U-u^{-1})\eta^\prime+[U-\frac{1}{2}(u^{-1}+u)]\eta=
\eta_1-\frac{1}{\omega}
\end{equation}
The general solution for the generating function $\eta(U)$ is
\begin{equation} \label{17}
\eta=I^{-1}[(\eta_1-\frac{1}{\omega})\int^U_0I^{-1}dU+C]
\end{equation}
where $I$, the integrating factor, is given for $U\leqslant u$ by 
\begin{equation} \label{18}
I=[(U-u)(U-u^{-1})^{1/2}=[1-\frac{2}{\omega}U+U^2]^{1/2}~.
\end{equation}
In general the expression (\ref{17}) has a factor behaving like
$(u-U)^{-1/2}$ which diverges at $U=u$. Its expansion as a power series
is of the form $u^{-1/2}\left(1+\frac{1}{2}u^{-1}U...+\frac{1/2~3/2
...\frac{2n-1}{2}}{n!}(u^{-1}U)^n + ...\right)$ which gives rise to
terms in $u^{-(n+1/2)}$ which diverge as $n$ becomes large because
$u<1$. To suppress this divergence we must choose the constant
\begin{equation*}
C=-\left(\eta_1 -\frac{1}{\omega}\right)\int^u_0I^{-1}dU~.
\end{equation*}
Then the numerator vanishes at $U=u$ and
\begin{equation*}
\eta=-\left(\eta_1-\frac{1}{\omega}\right)I^{-1}\int^u_U I^{-1}dU
\end{equation*}
But $\eta_0=\eta(0)=1$ so
$\eta_1=\frac{1}{\omega}-[\int^u_0I^{-1}dU]^{-1}$.

Now $I^2=\left[U-\frac{1}{2}(u^{-1}+u)\right]^2 -
\left[\frac{1}{2}(u^{-1}-u)\right]^2$ and $I(0)=1$, so we set
$\frac{1}{2}(u^{-1}+u)-U=\frac{1}{2}(u^{-1}-u)ch~w$; then
$I=\frac{1}{2}(u^{-1}-u)sh~w$ and $w=0$ when $U=u$. Also
\begin{equation*}
\int^u_U I^{-1}dU=w~.
\end{equation*}
Hence
\begin{equation} \label{19}
\eta(U)
=-\left[\frac{\eta_1-\frac{1}{2}(u^{-1}+u)}{\frac{1}{2}(u^{-1}-u)}\right]
~\frac{w}{shw}
\end{equation}
and
\begin{equation}\label{20}
\eta(0)=\left(\frac{1}{\omega}-\eta_1\right)ch^{-1}\gamma
\end{equation}
where
$\gamma^{-2}=1-\omega^2=1-v^2/c^2~;~\gamma=\frac{u^{2}+1}{u^{2}-1}$.
But $\eta(0)=\eta_0=1$ so we have
\begin{equation} \label{21}
\eta_1=1/\omega-1/ch^{-1}\gamma \rightarrow 1 ~{\rm
as}~\omega\rightarrow 1~.
\end{equation}
Thus at $\omega=1$ our recurrence relation (\ref{15}) yields the Magic
field with $\eta_n=1$ all $n$. We now know both $\eta_0=1$ and
$\eta_1(\omega)$ as a function of $\omega$; thus we are now in a
position to use our recurrence relation to generate all the $\eta_m$
for any chosen value of $\omega$. Alternatively we could expand our
expression for $\eta(U)$ as a power series in $U$ and pick out the
coefficient of $U^m$. However Prof J.F. Harper pointed out to me that
the recurrence relation (\ref{15}) is that obeyed by the Legendre
Polynomials $P_n\left(\frac{1}{\omega}\right)$. Notice that
$\frac{1}{\omega}$ is greater than 1. Since the $P_n$ and the $Q_n$
obey the same recurrence relation the general solution is a linear
combination of them. Fitting that combination to $\eta_0=1$ and the
value of $\eta_1$ just derived we find, noting that
$ch^{-1}\gamma\equiv Q_0\left(\frac{1}{\omega}\right)$,
\begin{equation} \label{22}
\eta_n=Q_n\left(\frac{1}{\omega}\right)/Q_0\left(\frac{1}{\omega}\right)~.
\end{equation}
Thus the solution for the complex potential is
\begin{equation} \label{23}
\Psi=\frac{q}{ia}\sum(2n+1)\!\left[Q_n
\!\left(\frac{1}{\omega}\right)\!/Q_0\!\left(\frac{1}{\omega}\right)\!\right]
\!Q_n\!\left(\frac{\tilde{r}}{ia}\right)\!P_n(\tilde{\mu})~.
\end{equation}

\vspace{5mm}
\noindent We note that
$Q_n\left(\frac{1}{\omega}\right)/Q_0\left(\frac{1}{\omega}\right)\rightarrow
1$ as $\omega \rightarrow 1$ so the Magic field is still there in that
limit. Now by [18] 8.8.2 $Q_n(z)\rightarrow
z^{-n-1}\int^\infty_0(1+cht)^{-n-1}dt$ as $z \rightarrow
\infty$. Hence
\begin{eqnarray*}
Q_n\left(\frac{1}{\omega}\right)/Q_0\left(\frac{1}{\omega}\right)&\rightarrow&
\omega^n\int^\infty_1
\frac{2^{n+1}x^n}{(1+x)^{2n+2}}dx \\
&=&
\begin{cases}
\omega^n2^n(n!)^2/(2n+1)!\\
\qquad + 0(w^{n+2})
\end{cases}
\end{eqnarray*}
so for small $\omega$
\begin{equation} \label{24}
\Psi=\frac{q}{ia}\sum\frac{2^n(n!)^2}{(2n)!}
\omega^nQ_n\left(\frac{\tilde{r}}{ia}\right) P_n(\bar{\mu})~.
\end{equation}

Now 
\begin{eqnarray*}
Q_0\left(\frac{\tilde{r}}{ia}\right) &=&
\ln\sqrt{\frac{\tilde{r}+ia}{\tilde{r}-ia}} =
i~\tan^{-1}\left(\frac{a}{\tilde{r}}\right)\\
Q_1\left(\frac{\tilde{r}}{ia}\right) &=& \frac{\tilde{r}}{a}
\tan^{-1}\frac{a}{\tilde{r}} -1~;\\
Q_2\left(\frac{\tilde{r}}{ia}\right)
&=&\left(\frac{-3\tilde{r}^2/a^2-1}{2}\right)
i~tan^{-1}\left(\frac{a}{\tilde{r}}\right) + \frac{i3\tilde{r}}{2a}
\end{eqnarray*}
so
\begin{eqnarray} \label{25}
\Psi&=&\frac{q}{a}\left\{\tan^{-1}\frac{a}{\tilde{r}}+i\omega
\left[1-\frac{\tilde{r}}{a}\tan^{-1}\left(\frac{a}{\tilde{r}}
\right)\right] \tilde{\mu}\right.\nonumber\\
&+&\left.\omega^2\left[\frac{\tilde{r}}{a}-\left(
\left(\frac{\tilde{r}}{a}\right)^2+3\right) \tan^{-1} \frac{a}{\tilde{r}}
\right]P_2(\tilde{\mu})\right.\nonumber\\
&+&\left.O(\omega^3)\right\}
\end{eqnarray}
The first term is the well known potential of a static charged disk.

We may now read off expressions for the charge $q$, the (magnetic)
dipole moment $\mu_m$ and the (electric) quadrupole moment $Q_e$
\begin{equation*}
\mu_m=qa\left(\frac{1}{\omega}-\frac{1}{\lambda}\right) \rightarrow 
\begin{cases}
\frac{1}{3}qa\omega +O(\omega^2)\\
qa~;~\omega \rightarrow 1
\end{cases}
\end{equation*}
where we have defined $\lambda
=\ln\sqrt{\frac{1+\omega}{1-\omega}}=ch^{-1}\gamma$.

In finding the quadrupole moment for $\omega$ small we note that part
of it comes from the oblateness of the $\tilde{r}=$const
surfaces. Taking the sign so an oblate distribution has negative $Q$
\begin{equation*}
Q_e=-\frac{1}{3}qa^2\!\left[1+2Q_2\left(\frac{1}{\omega}\right)/\lambda\right]
\!\rightarrow\!
\begin{cases}
-\frac{1}{3}qa^2\left(1+\frac{4}{5}\omega^2\right)\\
-qa^2~;~ \omega \rightarrow 1
\end{cases}
\!\!.
\end{equation*}

\section{Closed Form Exact Potential}

On the axis $\tilde{\mu}=1$, so all the $P_n(\bar{\mu})=1$ and we can
then perform the summation (\ref{23}) using a result in Whittaker \&
Watson [19]
\begin{eqnarray*}
\sum(2n&+&1)Q_n\left(\frac{1}{\omega}\right)Q_n
\left(\frac{\tilde{r}}{ia}\right)\\ &=&
\frac{1}{2\left(\frac{\tilde{r}}{ia}-\frac{1}{\omega}\right)} \ln
\left[\frac{1+\omega}{1-\omega}.\frac{\tilde{r}-ia}{\tilde{r}+ia}\right]
\end{eqnarray*}
On the axis $\tilde{r}=z$ so we obtain the complex potential setting
$b=\frac{a}{\omega}$
\begin{equation} \label{26}
\Psi(z) = \frac{q}{z-ib}\left[1+\frac{1}{2\lambda} \ln
\left(\frac{z-ia}{z+ia}\right) \right]~.
\end{equation}
Now there is a remarkable way [20] of obtaining the solution of
$\nabla^2\Psi=0$ everywhere if it is given as an analytic function
$\Psi(z)$ on the axis. The potential everywhere is given by (see
Appendix A)
\begin{equation*}
\Psi (R,z)=\frac{1}{2\pi}\int^{2\pi}_0\Psi(z+iR\cos\alpha)d\alpha~.
\end{equation*}

In our application it is simplest to use $\tan\frac{\alpha}{2}=t$ as
the variable of integration. The solution is then
\begin{equation*}
\Psi(R,z)=\frac{1}{\pi}\int^\infty_{-\infty}
\Psi\left(z+iR\frac{1-t^2}{1+t^2}\right) \frac{dt}{1+t^2}
\end{equation*}
applied to the $\Psi(z)$ given by (\ref{26}) we find
\begin{eqnarray} \label{27}
\Psi(R,z)=\frac{1}{\pi}\int^\infty_{-\infty}
\frac{q}{(z-ib+iR)+(z-ib-iR)t^2} \nonumber \\
\left\{1+\frac{1}{2\lambda}\ln
\left[\frac{(z-ia+iR)+(z-ia-iR)t^2}{(z+ia+iR)+(z+ia-iR)t^2}\right]\right\}dt~. 
\end{eqnarray}

Now
\begin{equation} \label{28}
\int^\infty_{-\infty}\frac{1}{g^2+t^2}=\frac{\pi}{g}
\end{equation}
and
\begin{equation} \label{29}
\int^\infty_{-\infty} \frac{\ln(f^2+t^2)}{g^2+t^2}dt = \frac{2\pi}{g}\ln(f+g)
\end{equation}
the latter result may be checked by differentiating with respect to
$f^2$ performing the integral that results by partial fractions and
finally reintegrating with respect to $f^2$. The constant of
integration is readily seen to be correct by taking the limit $f\gg g$;
for $z>0$ and $R>0$ no singularities occur in the $t$ integrations. We
are most interested in the region $R\leqslant a \leqslant b$ and $z$
small. We write 
\begin{eqnarray*}
g^2&=&\frac{z+iR-ib}{z-iR-ib} =\frac{b-R+iz}{b+R+iz}~;\\
f^2_+&=&\frac{z+iR-ia}{z-iR-ia}=\frac{a-R+iz}{a+R+iz} ~;\\
f^2_-&=&\frac{z+iR+ia}{z-iR+ia}=\frac{a+R-iz}{a-R-iz}~,
\end{eqnarray*}
and
\begin{equation*}
h^2=\frac{a+R+iz}{a-R-iz} =e^{i\pi}\frac{z-iR-ia}{z-iR+ia}~.
\end{equation*}
Writing $z=r\cos\theta,~R=r\sin\theta$ we see that for large
$r,~g,~f_+$ and $f_-$ all become $e^{i\theta}$ while $h$ is
$e^{i\pi/2}$. On the axis $R=0,~g,~f_+$ and $f_-$ are all unity while
$h$ is $e^{i\pi/2}\sqrt{\frac{z-ia}{z+ia}}$ which becomes 1 as
$z\rightarrow 0$. Performing the integrations and making sure that we
take the right branch at infinity we find
\begin{equation*}
\Psi=\frac{q}{(z-ib-iR)g}\left\{\!1+\frac{1}{\lambda}\!\left[\ln(-ih)
+\ln\!\left(\frac{f_++g}{f_-+g}\right)\right]\right\} ~,
\end{equation*}
so we have the solution to our problem
\begin{equation} \label{30}
\Psi=\frac{q}{\sqrt{R^2+(z-ib)^2}}\left[1+\frac{1}{\lambda}
Z-\frac{i\pi}{2\lambda}\right]~,
\end{equation}
Defining $S(a,R,z)=\sqrt{a+R+iz}$
\begin{equation} \label{31}
Z\!=\!\ln\!\!\left[\!\frac{S(a,-R,z) S(b,R,z)\!+\!S(a,R,z)  S(b,-R,z)}
                {S(a,R,-z)S(b,R,z)\!+\!S(a,-R,-z)S(b,-R,z)}\!\right]\!.
\end{equation}

\vspace{5mm}
\noindent Although this expression looks a bit formidable its derivatives at
$z=0$ where we shall need to evaluate them are quite simple
\begin{equation} \label{32}
\left.\frac{\partial Z}{\partial z}\right|_0 =
\frac{ib}{\sqrt{(a^2-R^2)(b^2-R^2)}}~,
\end{equation}
\begin{equation} \label{33}
\left.\frac{\partial Z}{\partial R}\right|_0
=\frac{R}{\sqrt{(a^2-R^2)(b^2-R^2)}}~.
\end{equation}
We have written the forms appropriate for $R\leqslant a$. 

In expressions (\ref{30}) and (\ref{31}) we have demonstrated how the
formidable sum (\ref{23}) is evaluated explicitly but we are not aware
that this identity has been demonstrated by experts in mathematical
functions.

It is now simple to check that the boundary condition (\ref{8}) which
may be rewritten 
\begin{equation*}
\left.\frac{\partial\Psi}{\partial
R}\right|_0=\left.\frac{iR}{b}\times\frac{\partial \Psi}{\partial
z}\right|_0~{\rm for}~R\leqslant a
\end{equation*}
is indeed satisfied by our solution (\ref{30}).  One consequence of
this is that it is easy to integrate the surface density of charge
\begin{equation*}
\sigma = \frac{1}{2\pi}\mathcal{R}e\left(-\frac{\partial\Psi}{\partial
z}\right)_0
\end{equation*}
evidently
\begin{eqnarray*}
\int^R_0 2\pi\sigma RdR&=&Re\int^R_0- \frac{b}{i} \frac{\partial
\Psi}{\partial R}dR \\
&=& - b \mathcal{I}m \left[\Psi
(R,0)-\Psi(0,0)\right]\\
Q(\leqslant R) &=& q\left\{1-\frac{b}{\sqrt{b^2-R^2}}\left[1+
\frac{Z(R,0)}{\lambda}\right]\right\}
\end{eqnarray*}
We note that $Z(a,0)=-\lambda$ because $b=a/\omega$ so as expected all
charge lies at $R\leqslant a$. As $\omega \rightarrow 1~Z\rightarrow
\ln \sqrt{1-\frac{R^2}{a^2}}$ for all $R<a$ but $\lambda \rightarrow
\infty$ 
\begin{equation*}
\lim_{\omega\rightarrow 1} \left\{\frac{Z(R,0)}{\lambda}\right\} =
\begin{cases}
~~0~{\rm for}~R<a\\
-1~{\rm for}~ R=a
\end{cases}
\end{equation*}
which explains the somewhat strange behaviour of $Q(<R)$ for the Magic
field.

At the pole the surface density of charge becomes zero when
$\lambda=\frac{1}{\omega}$ which occurs at $\omega=0.834$ which gives
a tip speed of $83.4\%$ of $c$. This should be compared with $93\%$
for a sphere. The difference is in the expected direction since the
disk has an uneven charge when $\omega=0$ and for it the Lorentz
magnetic force lies radially across the disk.

\section{Field Energy, Angular Momentum \& Gyromagnetic Ratio}

The total energy in the electromagnetic field is given by
\begin{eqnarray*}
\varepsilon&=&(8\pi)^{-1}\int(E^2+B^2)dV\\
&=& (8\pi)^{-1}\int\bm{\nabla}\Psi^*.\bm{\nabla}\Psi dV\\
&=&-(8\pi)^{-1}
\int^a_0\Psi^*\partial\Psi/\partial z 4\pi RdR\\
&=& i(b/2) \int^a_0\Psi^*\partial\Psi/\partial R dR\\
&=&-b/2~\mathcal{I}m
\int^a_0\Psi^*\frac{\partial\Psi}{\partial R}dR
\end{eqnarray*}
Now
\begin{eqnarray*}
\frac{\partial\Psi}{\partial
R}&=&\frac{qR^{3/2}}{(b^2-R^2)}\left[\frac{\pi}{2\lambda}
+i\left(1+\frac{Z}{\lambda}\right)\right]\\
&-&\frac{iqR}{\lambda
\sqrt{a^2-R^2}(b^2-R^2)}  
\end{eqnarray*}
and
\begin{equation*}
\Psi^*=\frac{q}{\sqrt{b^2-R^2}}
\left[\frac{\pi}{2\lambda}-i\left(1+\frac{Z}{\lambda}\right)\right]
\end{equation*}
so apart from the last term in $\partial\Psi/\partial R$, we see that
$\Psi^*$ and $\partial\Psi/\partial R$ have opposite phases. Thus only
the last term in $\partial\Psi/\partial R$ contributes to the
imaginary part of $\Psi^*\partial\Psi/\partial R$. Hence
\begin{equation*}
\varepsilon =\frac{\pi b}{8\lambda^2} q^2\int^a_0\frac{2RdR}{(a^2-R^2)^{1/2}(b^2-R^2)^{3/2}}~.
\end{equation*}
The integral is $2(a/b)(b^2-a^2)^{-1}$ so
\begin{equation} \label{34}
\varepsilon = \frac{\pi}{4\lambda^2}\frac{aq^2}{(b^2-a^2)}=
\pi(q^2/a)\frac{\omega^2}{(1-\omega^2)
\left[\ln\left(\frac{1+\omega}{1-\omega}\right)\right]^2}
\end{equation}
which gives the correct limit of $\frac{\pi}{4}(q^2/a)$ when $\omega
\rightarrow 0$.

The total angular momentum in the electromagnetic field is 
\begin{equation*}
{\bf{L}}=\frac{1}{4\pi c}\int {\bf{r}}\times
({\bf{E}}\times{\bf{B}})d V= \frac{1}{8\pi ic}\int
{\bf{r}}\times(\nabla\Psi^*\times \nabla\Psi )dV.
\end{equation*}
We note that the integral has no real part.

The Angular Momentum can be rewritten in the form

\begin{eqnarray*}
{\bf{L}}&=&
\frac{1}{4\pi c}
\int{\bf{r}}\times({\bf{E}}\times{\bf{B}})d^3x \\
&=&\frac{-1}{8\pi
c}{\mathcal{I}}m \int r\times [\nabla\Psi^*\times (\nabla AR \times
\nabla\phi)]d^3x\\
&=&\frac{1}{8\pi c} \mathcal{I}m \int {\bf{r}}\times \nabla\phi
\nabla\Psi^* . \nabla (AR)d^3x \\
&=& \frac{\hat{\bf{z}}}{8\pi c}
\mathcal{I}m \int \nabla\Psi^* . \nabla (AR)d^3x
\end{eqnarray*}
but since $\nabla^2\Psi^*=0$, div$(AR\nabla\Psi^*)
=\nabla\Psi^*.\nabla (AR)$ so we may turn {\bf L} into a surface
integral

\begin{equation*}
{\bf{L}}=\frac{-\hat{\bf{z}}}{8\pi c}\mathcal{I}m\int^a_0
AR\partial\Psi^*/\partial z 4 \pi RdR
\end{equation*}
where the factor 2 accounts for both sides of the disk. Now on the
disk $R\partial\Psi^*/\partial z= ib\partial\Psi^*/\partial R$ by
(\ref{8}) so on $z=0$,

\begin{eqnarray*}
{\bf{L}}&=&\frac{\hat{z}b}{2c}\mathcal{R}e\int^a_0
AR\partial\Psi^*/\partial R dR\\
&=&\frac{\hat{z}b}{2c}\mathcal{R}e\left[-\Psi^*(a,o)A(a,o)a +
\int^a_0\Psi^*\partial/\partial R (AR) dR\right]\\
&=&\frac{\hat{z}b}{2c}\mathcal{R}e\left[-\Psi^* Aa - \int^a_0\Psi^*
R\frac{\partial\Psi}{\partial z} dR\right]\\
&=&\frac{\hat{z}b}{2c}\left\{\mathcal{R}e\left[-a\Psi^*A\right]_a
-\mathcal{I}m[b\int^a_0\Psi^* \partial\Psi/\partial RdR]\right\}
\end{eqnarray*}

We already evaluated
$-\mathcal{I}m~b\int^a_0\Psi^*\partial\Psi/\partial RdR$ in connection
with the energy it gave $2\pi(q^2/a)\frac{\omega^2}{(1-\omega^2)
\left[\ln\left(\frac{1+\omega}{1-\omega}\right)\right]^2}$ so
${\bf{L}}=\pi{\bf{\hat{z}}}(q^2/c) \frac{\omega}{(1-\omega^2)\left[
\ln\left(\frac{1+\omega}{1-\omega}\right)\right]^2} -
\frac{\hat{\bf{z}}ab}{2c}\mathcal{R}e[\Psi^* A]_a$.  To evaluate the
last term we need to find $A$ from the formula (\ref{A2}) of the
Appendix putting $\tan\alpha/2=t$ this becomes
\begin{eqnarray*}
A(R,z)\!=\!\frac{iq}{\pi}\!\!\int\!\!\frac{(1-t^2)}
{(1+t^2)[(z-it+iR)+(z-ib-iR)t^2]}\\
\left\{ 1\!+\!\frac{1}{2\lambda} \ln
\left[ \frac{(z-ia+iR)+(z-ia-iR)t^2} {(z+ia+iR)+(z+ia-iR)t^2} \right]
\right\}dt
\end{eqnarray*}
Evidently we need the following integrals
\begin{eqnarray*}
I&=& \frac{1} {\pi} \int^\infty_{-\infty} \frac{1-t^2} {1+t^2}
\frac{\ln(f^2+t^2)} {g^2+t^2} dt \\
&=& 2 \left[ \frac{2} {(g^2-1)} \ln
\left( \frac{f+1} {f+g} \right) - \frac{(g-1)} {g(g+1)}\ln (f+g)
\right]
\end{eqnarray*}
This expression is derived by first differentiating $I$ with respect
to $f^2$ then integrating over $t$ and finally integrating with
respect to $f^2$. The final `constant' of integration which might
depend on $g$, is found to be zero by looking at the full limiting
values when $f\gg g$.
\begin{equation*}
K= \frac{1} {\pi} \int^\infty_{-\infty} \frac{1-t^2} {1+t^2} \frac{1}
{g^2+t^2} dt = - \frac{g-1} {g(g+1)}
\end{equation*}
Using these forms to evaluate $A$ we find

\begin{eqnarray*}
A(R,z)&=&\frac{iq}{(z-ib-iR)}
\left\{-\frac{g-1}{g(g+1)\lambda}\left[
\lambda - \frac{i\pi}{2} \right.\right.\\
&+&\left.\left.\ln h +
\ln\left(\frac{f_++g}{f_-+g}\right)\right] \right.\\
&+& \left. \frac{2}{(g^2-1)\lambda}
\ln \left[\left( \frac{f_++1} {f_-+1} \right) \left( \frac{f_-+g} {f_++g}
\right)\right] \right\}
\end{eqnarray*}
where $f_+,~f_-,~g$ and $h$ have the same definitions as before, and
$\ln h+\ln[(f_++g)/(f_-+g)]=Z$.

Although the above expression gives the complex $A$ everywhere, we
only need $A$ at $(a,o)$ to evaluate the angular momentum. There
$Z=-\lambda,~g=\sqrt{\frac{b-a}{b+a}} =
\sqrt{\frac{1-\omega}{1+\omega}},~f_+=0$ and $f_-\rightarrow\infty$
like $\sqrt{\frac{2a}{a-R}}$. With these substitutions the final $\ln$
term in $A$ becomes $\ln(1/g)=\lambda$.

Evaluating $A(a,o)$ we find

\begin{equation*}
A=\frac{q}{a}\left[\left(\frac{1}{\sqrt{1-\omega^2}}-1\right)
\frac{i\pi}{2} +1\right]
\end{equation*}
Now from (\ref{30}) $\Psi(a,o)=\frac{\pi
q\omega}{2a\sqrt{1-\omega^2}\lambda}$. Hence

\begin{equation*}
\mathcal{R}e(A\Psi^*)= \frac{\pi q^2}{2a^2}
\frac{\omega}{\sqrt{1-\omega^2}\lambda} 
\end{equation*}
and

\begin{equation*}
\frac{ab}{2c} \mathcal{R}e (A\Psi^*) =\frac{q^2}{4c}
\frac{1}{\sqrt{1-\omega^2}\lambda} 
\end{equation*}
I remind those who are worried that this may appear to have the wrong
behaviour with $\omega$ that
$\lambda=\ln\sqrt{\frac{1+\omega}{1-\omega}}$ is odd in $\omega$ and
$\simeq \omega$ for $\omega$ small. Putting this result into our
expression for $L$

\begin{equation} \label{35}
{\bf{L}}=\pi \hat{\bf{z}} \frac{q^2}{c} l(\omega)
\end{equation}
where

\begin{equation*}
l(\omega) = \frac{\omega-\frac{1}{2}\sqrt{1-\omega^2}
\ln\left(\frac{1+\omega}{l-\omega}\right)}{(1-\omega^2)\left[\ln\left(
\frac{1+\omega}{1-\omega}\right)\right]^2}
\end{equation*}
This expression tends to ${\bf{L}}=\pi
\frac{\hat{\bf{z}}q^2\omega}{6c}$ for small $\omega$. The
dimensionless quantity that is the inverse fine structure constant in
the case of the electron is

\begin{equation} \label{36}
2Lc/q^2 = 2\pi l(\omega)
\end{equation}
This assumes that all the angular momentum is electromagnetic. An
$\omega$ of 0.999743 gives the right value for the fine structure
constant; this corresponds to an equatorial Lorentz factor $\gamma$ of
44.1.

The gyromagnetic ratio is $\mu_m/L$ and the $g_m$ factor is
$\frac{2mc}{q} \frac{\mu_m}{L}$ so 

\begin{equation}\label{37}
g_m = 2\left(\frac{mc^2}{\varepsilon}\right) G(\omega)
\end{equation}
where $G(\omega)=
\frac{1-\omega/\lambda}{1-\sqrt{1-\omega^2}\lambda/\omega}$ with
$\lambda =\frac{1}{2}\ln\left(\frac{1+\omega}{1-\omega}\right)$. As
$\omega\rightarrow 1,~\lambda\rightarrow \infty$ but
$\sqrt{1-\omega^2}\lambda\rightarrow 0$, so $G\rightarrow 1$ and the
gyromagnetic ratio would tend to 2 if all the energy were
electromagnetic. However there is an unbalanced electromagnetic stress
on the disk's edge and most ways of making a balancing stress also add
to the energy. Some add to the angular momentum too. Perhaps the
simplest way to a consistent model is to add a string loop around the
edge of the disk with energy per unit length equal to its tension
$T$. Such a string loop carries no angular momentum but has energy
$2\pi aT$. The total energy of the whole configuration is then
$\varepsilon+2\pi aT$. Minimising this over `$a$' while keeping the
angular momentum $L$ fixed via (\ref{36}) means keeping $\omega$
fixed. Hence from (\ref{34}) $\varepsilon$ is proportional to $a^{-1}$
and equilibrium at the minimum is achieved when $2\pi
Ta=\varepsilon$. Thus the total energy is $2\varepsilon$ and mass of
the whole configuration at equilibrium will be $2\varepsilon/c^2$; for
such a model the gyromagnetic ratio or rather the $g_m$ factor is
\begin{equation*}
g_m=4G(\omega)
\end{equation*}
which becomes not 2 but 3.46 for the $\omega$ that gives the correct
fine structure constant. Probably it is more sense to model the disk
with a stress-energy tensor based on the Kerr disk (Lasenby {\it et
al.} 2004 [21]) which contains both angular momentum and energy.

It is sensible to remark that infinities such as that found here as
$v\rightarrow c$ are often removed when relativistic problems are
treated quantum mechanically.

In [3] we gave singular solutions that rotated uniformly at any rate
and were superpositions of forward and backwardly rotating Magic
fields. The equal superposition was static but did {\it not} give the
well known solution of the charged static disk. This gave us doubts as
to whether the Magic field would be the $v\rightarrow c$ limit of the
rotating disk problem but we have now shown it to be so. The other
singular solutions are discussed in Appendix B.

\begin{acknowledgments}
I thank Prof J.F. Harper, my brother-in-law, for his mathematical
help. Some of this work was done at the Institute of Advanced Study
and the support of the Monell Foundation is gratefully
acknowledged. Dr N.W. Evans was helpful in discussing calculational
details.
\end{acknowledgments}

{}

\appendix
\section{Scalar and Vector Potentials for Poloidal Fields}

We consider fields which lie in the meridian planes through an axis of
symmetry and which are harmonic above the plane $z=0$. For simplicity
of exposition we take their potentials to be
O$\left(\frac{1}{r}\right)$ at $\infty$ but the method is readily
extendable beyond those cases. Since the fields are harmonic we may
write
\begin{equation*}
{\bf{E}}+i{\bf{B}}=-\nabla\Psi=~{\rm Curl}~{\bf{A}}
\end{equation*}
It is normally not hard to derive an expression for $\Psi$ on the axis
of symmetry and we shall suppose this has been done so that $\Psi(z)$
is known, and $\Psi(z)\rightarrow q/z$ as $z\rightarrow \infty$.

We show that the complex scalar and vector potentials $\Psi(R,z)$ and
$A\hat{\phi}$ where $\hat{\phi}$ is the unit toroidal vector are given by
\begin{equation} \label{A1}
\Psi(R,z)=\frac{1}{2\pi}\int^{2\pi}_0\Psi (z+iR\cos\alpha )d\alpha
\end{equation}
\begin{equation} \label{A2}
A(R,z)= \frac{i}{2\pi} \int^{2\pi}_0 \cos\alpha\Psi (z+iR\cos\alpha )d\alpha
\end{equation}
provided $\Psi(z)$ is analytic in $z<0$.

Proof:
\begin{equation*}
\Psi (z) = \frac{1}{2\pi i} \oint \frac{\Psi(\zeta)}{(\zeta
-z)}d\zeta
\end{equation*}
where the path in the complex $\zeta$ plane is chosen to encircle
$\zeta =z$.

\noindent Hence 
\begin{equation*}
\Psi (z+iR\cos\alpha)=\frac{1}{2\pi i} \oint \frac{\Psi (\zeta)}{\zeta
-z-iR\cos\alpha}d\zeta
\end{equation*}
and 
\begin{equation*}
\int\!\Psi (z+iR\cos\alpha)d\alpha = \frac{1}{2\pi i} \int^{2\pi}_0
\!\oint \frac{\Psi (\zeta)}{\zeta -z-iR\cos\alpha}d\zeta d\alpha
\end{equation*}
We choose the integration circle to surround the whole circle
$z+iR\cos\alpha$ for all $\alpha$. We write $\tan \alpha/2 = t$ and
reverse the order or integration
\begin{eqnarray*}
\int^\infty_{-\infty} \frac{dt}{(\zeta -z-iR)+(\zeta
-z+iR)t^2}&=& \frac{\pi} {\sqrt{(\zeta -z)^2+R^2}}\\
&=& \frac{\pi}{\sqrt{({\bf{r}}-\zeta\hat{\bf{r}})^2}}
\end{eqnarray*}

Thus
\begin{equation*}
\frac{1}{2\pi} \int^{2\pi}_0 \Psi (z-iR\cos\alpha)d\alpha =
\frac{1}{2\pi i} \oint \frac{\Psi
(\zeta)}{\sqrt{({\bf{r}}-\zeta\hat{\bf{r}})^2}} 
\end{equation*}
But $\nabla^2 \frac{1}{\sqrt{({\bf{r}}-\zeta \hat{\bf{r}})^2}}$ is zero
everywhere except at ${\bf{r}}=\zeta\hat{\bf{r}}$ which is not
encountered on the integration path.  Hence
$\frac{1}{2\pi}\int^{2\pi}_0 \Psi (z+iR\cos\alpha)dx$ is harmonic. But
it also reduces to $\Psi(z)$ on axis and is
$O\left(\frac{1}{r}\right)$ at $\infty$ so it gives the complex
potential everywhere $z\geqslant 0$. 

To obtain the vector potential we note that
\begin{equation} \label{A3}
{\rm Curl}{\bf{A}}=\nabla (AR)\times\nabla\phi =-\nabla\Psi
\end{equation}
and
\begin{equation*}
{\rm Curl~Curl}A=-[\nabla^2(AR)-2\nabla\ln R\nabla
(AR)]\nabla\phi=0
\end{equation*}
for $z>0$ because there are no sources above $z=0$.

We now write $A=\partial\zeta/\partial R$ then the expression in
square brackets above reduces to 
\begin{equation*}
R~ \partial/\partial R (\nabla^2\zeta)=0
\end{equation*}
so
\begin{equation*}
\nabla^2\zeta =f(z)
\end{equation*}

Now we may add an arbitrary function of $z$ to $\zeta$ without
changing $A$ so we add $-\iint^zf(z)dz$ and deduce that the new
$\zeta$ will obey
\begin{equation*}
\nabla^2\zeta=0
\end{equation*}
from (\ref{A3}) and the above
\begin{equation*}
-\frac{\partial^2\zeta}{\partial z^2} = \frac{1}{R}
 \frac{\partial}{\partial R} \left(R\frac{\partial\zeta}{\partial R}\right)
 =-\frac{\partial\Psi}{\partial z}
\end{equation*}
Hence $\zeta (z)=\int^z \Psi(z^\prime)dz^\prime$.

The linear function of $z$ that comes from the integration can again
be absorbed into $\zeta$ without changing $A$ or the fact that
$\nabla^2\zeta=0$.

We now use the theorem proved above for taking $\Psi$ off axis, but
apply it to $\zeta(z)$. Then
\begin{equation*}
\zeta (R,z)=\frac{1}{2\pi}\int^{2\pi}_0 \int^{z+iR\cos\alpha}\Psi
(z^\prime)dz^\prime d\alpha
\end{equation*}
differentiating with respect to $R$
\begin{equation*}
A=\frac{i}{2\pi}\int^{2\pi}_0\cos\alpha\Psi(z+iR\cos\alpha )dx\alpha~.
\end{equation*}

We notice that both $\zeta$ and $A$ obey their appropriate equations
and that $A$ is $O\left(\frac{1}{r}\right)$ at $\infty$ as required.

\vspace{5mm}
\section{Singular Solutions}

In [3] we remarked that if $\Psi=q[R^2+(z-ia)^2]^{-1/2}$ is the
complex potential of the Magic field itself then
$\alpha\Psi+(1-\alpha)\Psi^*$ is a superposition of a forward rotating
and a backward rotating Magic field. The electrical potential of the
disk is still zero and no magnetic field crosses it in $R<a$. There is
the same charge density on the disk as in $\Psi$ and it rotates
uniformly but less fast. When $\alpha=\frac{1}{2}$ the charge does not
rotate but the surface density is still singular. This is {\bf not}
the solution for the non-rotating charged disk but to what physical
problem does it correspond? I believe it must correspond to the field
of an earthed plate surrounded by a charged wire with a small
insulating gap, between the wire and the plate. As the gap becomes
smaller, a larger and larger charge must be placed on the wire to
ensure that the net charge of wire and plate remains at a fixed value
$q$. I think this non-rotating magic field corresponds to such a
singular limit. Perhaps the rotating versions with $\alpha \neq
\frac{1}{2}$ correspond to the singular solutions of the recurrence
relations that we rejected. These solutions obey the conditions
required of a superconducting disk, the charges while rotating
uniformly do not correspond to the physical problem that we set out to
solve. Nevertheless the Magic field itself is the $v\rightarrow c$
limit of our problem.

\end{document}